\documentclass[12pt]{article}
\usepackage{amssymb,amscd,amsmath}

\newcommand{\2}{{\hat sl}_2}
\newcommand{\1}{{\bf CP}^1}
\newcommand{\pf}{\Phi^{j}}
\newcommand{\bx}{\bar x}
\newcommand{\sx}{x^{\ast}}

\newcommand{\gm}{\Gamma}
\newcommand{\bm}{j_{-}}
\newcommand{\bp}{j_{+}}
\def\n#1{{}_{#1}}

\def\pf#1#2{\Phi^{#1\,#2}}
\def\Q{\mathbb{Q}}
\def\N{\mathbb{N}}
\def\h{\frac{1}{2}}
\def\NP#1#2{{\it Nucl.Phys.} {\bf B#1} (#2)}
\def\PL#1#2{{\it Phys.Lett.} {\bf B#1} (#2)}

\def\MPL#1#2{{\it Mod.Phys.Lett.} {\bf A#1} (#2)}
\title{On 2D gravity coupled to c$\leq$ 1 matter in Polyakov
light-cone gauge}
\author{Oleg Andreev\thanks{e-mail: andre@landau.ac.ru} \\ \\
L.D.Landau Institute for Theoretical Physics,
\\ Kosygina 2, 117940 Moscow, Russia}
\date{}
\begin{document}
\maketitle

\begin{abstract}
A new formulation of $c\leq 1$ matter coupled to 2D gravity is proposed.
This model, being closely analogous to one in the Polyakov light-cone gauge,
possesses well defined global properties which allow to calculate correlation
functions. As an example, the three point correlation functions of discrete
states are found.
\end{abstract}

\vspace{-11cm}
\hfill LANDAU-96-TMP-1

\hfill hep-th/9601026

\vspace{11.5cm}

Since the seminal works of Polyakov, Knizhnik and Zamolodchikov\cite{P1,KPZ},
there has been much progress in understanding the continuum field theory
approach to 2D gravity (see e.g.\cite{AD} and refs. therein).
Majority of efforts has been devoted to the study of the coupling of conformal
matter to gravity in the conformal gauge. The reason why it is useful
lies in the facts that it is the standard gauge for conformal field theory
and its properties on the Riemann surfaces are well known. At the same time,
the properties of the Polyakov gauge are little known which restricts the
applications of such a gauge. In this letter I will
present results of a new formulation of c $\leq$ 1 matter coupled to 2D
gravity motivated by the hope that this theory, being closely analogous to one
in the Polyakov gauge, possesses well defined global properties which permit to
calculate correlation functions.

The starting point is the G/G topological model for G=SL(2) \cite{GK}.
The Hilbert space of the model decomposes into holomorphic and
anti-holomorphic sectors. For my purposes, I need only one of them.
Let me choose the holomorphic sector. It has $\2\oplus\2\oplus\2$
as the symmetry algebra. The corresponding currents form the following OP
algebras
\begin{align}
J^{\alpha}_a(z_1)J^{\beta}_a(z_2)=\frac{k_a}{2}q^{\alpha\beta}\frac{1}{(z_1-z_2)^2}+
\frac{f^{\alpha\beta}_{\quad\gamma}}{(z_1-z_2)}J^{\gamma}_a(z_2)+O(1)\qquad,
\end{align}
where $q^{\alpha\beta}$ is the Killing metric of $sl_2$,  $
f^{\alpha\beta}_{\quad\gamma}$ are its structure constants and $a$ is a
parameter labelling algebras in the direct sum. One can always choose a basis
of $sl_2$ such that
$ q^{00}=1,\,\,q^{+-}=q^{-+}=2,\,\,f^{0+}_{\quad +}=f^{-0}_{\quad -}=1,\,\,
f^{+-}_{\quad 0}=2;\,\,\alpha,\beta,\gamma=0,+,-$. The levels are given by
\begin{align}
k_1=k\,\,,\qquad k_2=-k-4\,\,,\qquad k_3=4\,\,.
\end{align}
It should be noted that $J^{\alpha}_3$ are expressed in terms of the first
order femionic systems(ghosts) of weights (1.0). Explicitly $J^{\alpha}_3(z)=
f^{\alpha}_{\,\,\,\beta\gamma}:b^{\beta}(z)c^{\gamma}(z):.$ The Fock space
$\cal F {}^{\text{gh}}$ is built on a vacuum state
$|0\rangle{}^{\text{gh}}$ by acting with the corresponding modes of the free
fields\footnote{Note that in my conventions
$|0\rangle{}^{\text{gh}}=c^0_0|0\rangle\otimes c^+_0|0\rangle\otimes c^-_0
|0\rangle$ .}.
Its structure is described in terms of the irreducible representations of
$\2$, namely it is a direct sum of the integrable representations.
Normalize the ghost number such that the vacuum $|0\rangle{}^{\text{gh}}$ has
ghost number zero. I will focus on physical operators at
ghost number zero so I don't explicitly write down the dependence on ghosts in
most cases in what follows.
\newline The stress-energy tensor is a sum of
Sugawara terms of $J^{\alpha}_1$ and $J^{\alpha}_2$ currents and the usual
contribution of the ghost systems:
\begin{align}
T(z)=\frac{1}{k+2}q_{\alpha\beta}:J^{\alpha}_1(z)J^{\beta}_1(z):-
\frac{1}{k+2}q_{\alpha\beta}:J^{\alpha}_2(z)J^{\beta}_2(z):+q_{\alpha\beta}
:b^{\alpha}(z)\partial c^{\beta}(z):\,\,.
\end{align}
It is easy to see that the Virasoro algebra generated by this stress-energy
tensor has zero central charge.
\newline As to the algebraic structure and physical states of the model I
refer to refs.[4,7].

In general, when
given representations of a chiral algebra(symmetry algebra), in order to
define fields of quantum field theory, one needs a construction attaching
representations to a point. In [5] Feigin and Malikov proposed the improved
construction for the case of $\2$
(see also \cite{FZ}). The point is that a module should be
attached to a pair $(x,z)$. The coordinate $z$ is a point on a curve.
As to $x$, it must be taken as a point on $\1$. Note that in physics $x$ is
called as isotopic coordinate.  The generators of $sl_2$ are given by
\begin{align} S_j^{-}=\frac{\partial}{\partial x}\,,\qquad
S_j^{0}= -x\frac{\partial}{\partial x}+j\,,\qquad
S_j^{+}=-x^2\frac{\partial}{\partial x}+2jx\,.
\end{align}
Here $j$ is the weight of representation.
\newline The chiral currents are turned into a form(current)
\begin{align}
J(x,z)=J^+(z)-2xJ^0(z)-x^2J^-(z)\qquad.
\end{align}
It is easily shown that the Operator Product (OP) expansion of $J(x,z)$ is
\begin{align}
J(x_1,z_1)J(x_2,z_2)=-k\frac{x_{12}^2}{z_{12}^2}-2\frac{x_{12}}{z_{12}}
J(x_2,z_2)-\frac{x_{12}^2}{z_{12}}\frac{\partial}{\partial x_2}J(x_2,z_2)+O(1)
\qquad,
\end{align}
where $z_{ij}=z_i-z_j\,,\,\,\, x_{ij}=x_i-x_j\,$.
\newline The primary fields are defined via their OP expansions with the
current
\begin{align}
J(x_1,z_1)\pf{j}{}(x_2,z_2)=-2j\frac{x_{12}}{z_{12}}
\pf{j}{}(x_2,z_2)-\frac{x_{12}^2}{z_{12}}\frac{\partial}{\partial x_2}
\pf{j}{}(x_2,z_2)+O(1)\quad.
\end{align}
\newline It should be noted that in the general case the primary fields are
non-polynomial in $x$. Furthermore, $J(x,z)$ is not primary.

It is now straightforward to use this machinery in the case at hand. Let
$J_1(x,z)$ and $J_2(\bx,z)$ be the corresponding forms of the algebras in the
direct sum\footnote{In fact, in the case of integer levels one doesn't need
the $x$ variable, so
the ghosts don't lead to an additional isotopic coordinate.
}. The primary fields at ghost number zero are given by
\begin{align}
\pf{j_1,}{j_2}(x,\bx,z)=\pf{j_1}{}(x,z)\pf{}{j_2}(\bx,z)\quad.
\end{align}
Here $\pf{j_1}{}(x,z)(\pf{}{j_2})$ is primary with respect to $J_1(x,z)(J_2)$.

The idea that the SL(2)/SL(2) model is connected to the minimal models
coupled to gravity was put forward in ref.\cite{Aetal}, in a study of some
"numerological" correspondences and partition functions. This discusses
mainly the conformal gauge.

Let me now clarify some points in my framework.
Setting $x=\bx=z$, which corresponds to the quantum hamiltonian
reduction of $\2\oplus\2$ to $Vir\oplus Vir$ \cite{PG}, one immediately
obtains the minimal model coupled to gravity, more correctly its holomorphic
sector in the conformal gauge \footnote{Note that $Vir$  means
the Virasoro algebra.}. In this case the first $Vir$ describes the matter
sector. The second $Vir$ corresponds to the Liouville(gravity) sector.
It is straightforward to see that, under $x=\bx=z$, $J^-_1(z)$ and $J^-_2(z)$
are constrained. It leads to the following stress-energy tensors
\begin{align}
T_a(z)=\frac{1}{k+2}q_{\alpha\beta}:J^{\alpha}_a(z)J^{\beta}_a(z):-\partial
J^0_a(z)\qquad.
\end{align}
In terms of fields the reduction is given by
\begin{align}
\pf{j_{n.m}}{}(x,z)\pf{}{-1-j_{n.m}}(\bx,z)\rightarrow
\phi_{n.m}(z){\text{exp}}\beta_{n.m}\varphi(z)\qquad .
\end{align}
In the above, $j_{n.m}$ take values defined by the Kac-Kazhdan list \cite{KK}.
Namely
\begin{align}
j^-_{n.m}=\frac{n-1}{2}\bm+\frac{m-1}{2}\bp\,\qquad\text{ or }\qquad
j^+_{n.m}=-\frac{n+1}{2}\bm-\frac{m}{2}\bp\quad,
\end{align}
with $\bm=1\,,\,\bp=-k-2\,,\,k\in\Q\,,\,\{n,m\}\in\N$ \footnote
{For the rational level k the weights given in (11) are called admissible
\cite{KW}.}.  \newline As to the right-hand side it is the primary field
$\phi_{n.m}(z)$ of the minimal conformal theory dressed by the Liouville
exponent (see e.g. \cite{AD} and refs. therein).

It is surprising that there exists another construction which represents an
analog of the minimal conformal matter coupled to gravity in the Polyakov
gauge. Let me explain how this idea can be implemented.
In contrast to the previous case set $x=z$.
>From this it follows that only $J_1^-(z)$ is constrained.
As a result one has $Vir\oplus\2$ as the symmetry algebra. The
stress-energy tensors are those given in (9). It is worth noting that
they take such form due to entirely different reasons, namely the quantum
hamiltonian reduction and decomposition (5) respectively.

For the primary fields one obtains
\begin{align}
\pf{j_{n.m}}{}(x,z)\pf{}{j_{n.m}}(\bx,z)\rightarrow
\phi_{n.m}(x)\pf{}{j_{n.m}}(\bx,x)\qquad ,
\end{align}
where $j_{n.m}$'s are from the Kac-Kazhdan list. It should be stressed that
$\pf{}{j_{n.m}}(\bx,x)$ is primary with respect to $\2$ but not with respect
to $T$ given by eq.(9).

Now let me show that the proposed construction provides all features
of the minimal models coupled to 2D gravity in the Polyakov gauge.

It is easy to check that a condition $c^{\text{tot}}=0$
is equivalent to a relation for the
levels $k_1+2=-k_2-2$ given by eq.(2). The same is also true for the
conformal gauge where it automatically leads to a relation between background
charges of the matter and Liouville sectors \cite{AD}.

The KPZ scaling law \cite{P1,KPZ}, determining the $\2$ weights of the
primary (spinless) field $\phi_{n.m}$ interacting with gravity is satisfied
by setting $j_1=j_2$ for the primary fields (8). By the
way, in the case of the conformal gauge a proper Liouville exponent
is reproduced by $j_1=-j_2-1$.

Moreover the residual $\2$ algebra assumes the Knizhnik-Zamolodchikov (KZ)
equation for the correlators of the primary fields $\pf{}{j_{n.m}}(\bx,x)$.
Explicitly
\begin{align}
-(k+2)\frac{\partial}{\partial x_i}\langle \pf{}{j_1}(\bx_1,x_1)\dots
\pf{}{j_N}(\bx_N,x_N)\rangle=\sum^{N}_{i\not=l}\frac{q_{\alpha\beta}
S^{\alpha}_{j_i}S^{\beta}_{j_l}}{x_i-x_l}
\langle \pf{}{j_1}(\bx_1,x_1)\dots\pf{}{j_N}(\bx_N,x_N)\rangle \,,
\end{align}
where $S^{\alpha}_{j_i}$ are the generators of $sl_2$ (4) i.e. the
differential operators with respect to $\bx_i$. Note that
the term $\partial J_2^0(x)$ modifying the stress-energy tensor doesn't affect
the KZ equation because the current $J_2^0(x)$ has no $\log x$ terms in
its mode expansion.

In contrast to the Polyakov gauge where a global structure of 2d world sheet
is unclear in the case at hand one has a well-defined ${\bf CP}^1\times{\bf
CP}^1$ structure. It allows to solve the KZ equation for the admissible
representations by the methods of conformal field theory (see for
instance \cite{A}).

Of course in the above I have not said anything specific about the
BRST analysis of physical states. In order to find them one must
solve the BRST cohomology problem. I refer to the paper by Marcus and Oz
 for more details \cite{MO}.

Now let me give an explicit calculation of correlation functions.
My aim is to find the three point functions of operators
\begin{align}
{\mathcal O}\n{n.m}=
\int d\mu_{n.m}(x,\bx;k)\phi_{n.m}(x)\pf{}{j_{n.m}}(\bx,x)\quad .
\end{align}
Here
$\mu_{n.m}(x,\bx;k)$ represents a measure which will be defined later.
$\phi_{n.m}(x)\pf{}{j_{n.m}}$ are the primary fields (12).
Having set notations as above one gets
\begin{align}
\langle{\mathcal O}\n{n_1.m_1}{\mathcal O}\n{n_2.m_2}{\mathcal O}\n{n_3.m_3}
\rangle=\langle\prod_{i=1}^3\int d\mu_{n_i.m_i}(x_i,\bx_i;k)
\phi_{n_i.m_i}(x_i)\pf{}{j_{n_i.m_i}}(\bx_i,x_i)\rangle\quad .
\end{align}
The integrand is factorized into two pieces:
\begin{align}
\langle
\phi_{n_1.m_1}(x_1)\phi_{n_2.m_2}(x_2)\phi_{n_3.m_3}(x_3)\rangle
\langle\pf{}{j_{n_1.m_1}}(\bx_1,x_1)\pf{}{j_{n_2.m_2}}(\bx_2,x_2)
\pf{}{j_{n_3.m_3}}(\bx_3,x_3)\rangle\quad .
\end{align}
These correlators are standard, and I find
\begin{align}
\langle\phi_{n_1.m_1}(x_1)\phi_{n_2.m_2}(x_2)\phi_{n_3.m_3}(x_3)\rangle=
C_{(n_2.m_2)(n_3.m_3)}^{(n_1.m_1)}\,
\prod_{i<l}\frac{1}{(x_{il})^{\gamma_{il}
({\Delta}_0)}}\quad,
\end{align}
\begin{align}
\langle\pf{}{j_{n_1.m_1}}(\bx_1,x_1)\pf{}{j_{n_2.m_2}}(\bx_2,x_2)
\pf{}{j_{n_3.m_3}}(\bx_3,x_3)\rangle=
\Tilde C_{(n_2.m_2)(n_3.m_3)}^{(n_1.m_1)}\,
\prod_{i<l}\frac{(\bx_{il})^{\gamma_{il}(j)}}{(x_{il})^{\gamma_{il}
(\Delta)}}\quad,
\end{align}
with
$y_{nm}=y_n-y_m\,,\,\gamma_{12}(y)=y_1+y_2-y_3\,,\,
\gamma_{13}(y)=y_1+y_3-y_2\,,\, \gamma_{23}(y)=y_2+y_3-y_1\,$
and
\begin{align*}
{\Delta}_0=\frac{j(j+1)}{k+2}-j \,\,,\qquad  \qquad
\Delta=-\frac{j(j+1)}{k+2}\quad.
\end{align*}
Moreover, $C(\Tilde C)$ are the square
roots of the structure constants for the minimal models and SL(2) CFT
respectively.

Substituting (17) and (18) into (15) one arrives at
\begin{align}
\begin{split}
\langle{\mathcal O}\n{n_1.m_1}{\mathcal O}\n{n_2.m_2}{\mathcal O}\n{n_3.m_3}
\rangle&=
C_{(n_2.m_2)(n_3.m_3)}^{(n_1.m_1)}\,
\Tilde C_{(n_2.m_2)(n_3.m_3)}^{(n_1.m_1)}\,  \\
&\times\prod_{i=1}^3\int d\mu_{n_i.m_i}(x_i,\bx_i;k)
\prod_{i<l}(\bx_{il})^{\gamma_{il}(j)}(x_{il})^{\gamma_{il}(j)}\quad.
\end{split}
\end{align}

Let me now consider the euclidian domain $\bx_i=\sx_i$, where the star
denotes the complex conjugation. The correlation function is rewritten as
\begin{align}
\langle{\mathcal O}\n{n_1.m_1}{\mathcal O}\n{n_2.m_2}{\mathcal O}\n{n_3.m_3}
\rangle=
C_{(n_2.m_2)(n_3.m_3)}^{(n_1.m_1)}\,
\Tilde C_{(n_2.m_2)(n_3.m_3)}^{(n_1.m_1)}
\,\prod_{i=1}^3\int d\mu_{n_i.m_i}(x_i,\sx_i;k)
\prod_{i<l}\vert x_{il}\vert^{2\gamma_{il}(j)}\,\,\,.
\end{align}

The factor $C\Tilde C$ can be found from the explicit expressions of
the structure constants \cite{DF,A}, after some simple but tedious algebra.
Unfortunately this is not the case for the integral at generic weights
$j_{n.m}$. However, if $m$ is equal to 1, then $j^+_{n.1}$ is an integer or
half-integer. At these values of the weights the primary fields form SU(2)
multiplets \cite{FZ}. The integrand is the generating function for the
Clebsch-Gordan coefficients of SU(2). As to the measure, one can consider the
limit $k\rightarrow\infty$. In this limit it is the standard SU(2) invariant
measure. Explicitly

\begin{align} d\mu_n(x,\sx)=\frac{d^2x}{(1+\vert x\vert
^2)^{n+1}}\quad.
\end{align}

The integral in eq.(20) reduces to
\begin{align}
I(n_1,n_2,n_3)=
\prod_{i=1}^3\int\,\frac{d^2x_i}{(1+\vert x_i\vert ^2)^{n_i+1}}
\prod_{i<l}\vert x_{il}\vert^{\gamma_{il}(n-1)}\quad.
\end{align}
It assumes a remarkably simple form
\begin{align}
I(n_1,n_2,n_3)=\gm(\sigma+\h)\prod_{i=1}^3\,\frac{\gm(\sigma-n_i+\h)}
{\gm(n_i+1)} \quad,
\end{align}
where $\sigma=\frac{n_1}{2}+\frac{n_2}{2}+\frac{n_3}{2}$.
\newline The integral has been calculated for some cases, by using the
following normalization $\int d\mu_n(x,\bx)\vert
x\vert^{2m}=\gm(m+1)\gm(n-m)/\gm(n+1)$, and then the general
form (22) has been guessed.

Using the expressions for the structure constants and result (23) the three
point function of the operators (14) with $j_{n.m}=j^-_{n.1}$ can be found in
the form
\begin{align}
\langle{\mathcal O}^-_{n_1.1}{\mathcal
O}^-_{n_2.1}{\mathcal O}^-_{n_3.1}
\rangle=\Bigl(\frac{1}{n_1n_2n_3}\Bigr)^{\h}\quad.
\end{align}
The non-trivial $n$-dependence cancels out, and I end up only with leg
factors.

Finally, normalize the correlation functions in the same way as in
\cite{DKetal} one gets
\begin{align}
\langle{\mathcal O}^-_{n_1.1}{\mathcal
O}^-_{n_2.1}{\mathcal O}^-_{n_3.1} \rangle_{\text {norm}}=n_1 n_2 n_3\quad.
\end{align}
This formula agrees with both the matrix model result and conformal gauge one
\cite{DKetal,AD}. Note that the operators ${\mathcal O}^-_{n.1}$
correspond to $\Phi^-_{n.1}$ operators in the conformal gauge. Thus they
represent discrete states in the Polyakov light-cone gauge.

To summarize, the main point in this letter is the well defined structure of
2d world sheet. It allows to avoid a question on a global fixing of the
Polyakov gauge. Moreover all properties of $c\leq 1$ matter coupled to 2D
gravity in the Polyakov gauge are retained. So one gets better control of
the model. The construction reminds one of an idea by Schwarz \cite{S} that
there exists a well defined gauge so that the theory has the same properties
as in the Polyakov gauge.

Let me also mention some interesting features of the construction together
with open problems.
\newline (i) One is on a "world sheet-isotopic" transmutation. Indeed,
starting with the SL(2)/SL(2) model, and defining $z$ as the world sheet
coordinate and $x,\bx$ as the isotopic ones, one arrives at a rather amusing
picture:  $x,\bx$ become the world sheet coordinates of the model.
\newline (ii) The chiral sector of the SL(2)/SL(2) theory reduces to the
chiral sector of the minimal models coupled to gravity in the conformal gauge
under the quantum hamiltonian reduction. On the other hand it is possible
to reduce the same sector to the full theory for the Polyakov gauge. So,
one can imagine that this gauge provides a "minimal" description of the model.
\newline (iii) In order to calculate the correlation functions for
${\mathcal O}\n{n.m}$ operators one needs SL(2) invariant measure $d\mu$
which depends on $k$ in a rather nontrivial way as well as integrands.
In fact integrands for the four point (etc.) correlation functions are known
only for the simplest case of the free fermions where they were found due to
the path integral methods \cite{BK}. Unlike the minimal models there is no
general principle for combining the conformal blocks in the model. The
obvious origin of this trouble is that the number of conformal blocks in the
minimal models and SL(2) conformal field theory are different. The problem is
to find measure and integrands.

I am grateful to B.Feigin, R.Metsaev and A.S.Schwarz for useful
discussions and G.Lopes Cardoso for reading the manuscript. I would also like
to thank M.Lashkevich  and N.Marcus for drawing my attention to
refs.\cite{Aetal} and \cite{MO} respectively. This work was supported in
part by INTAS grant 94-4720.


\begin{thebibliography}{99}

\bibitem{P1}
A.Polyakov, \MPL{2}{1987} 893;
in Les Houches 1988: Two-dimensional quantum gravity. Superconductivity at high
$T_c$.

\bibitem{KPZ}
V.Knizhnik, A.Polyakov and A.Zamolodchikov, \MPL{3}{1988} 819.

\bibitem{AD}
J.Ambj\o rn, in Les Houches 1990: Quantization of Geometry, Preprint
NBI-HE-94-53; \\
L.Alvarez-Gaume and C.Gomez, Topics in Liouville Theory, Lectures at
the Trieste Spring School, 1991, Preprint CERN-TH.6175/91; \\
F.David, in Les Houches 1992: Simplicial Quantum Gravity and Random Lattices,
Preprint Saclay T93/028.

\bibitem{GK}
K.Gawedzki and A.Kupianen, \PL{215}{1988} 119, \NP{320}{1989} 649.

\bibitem{FM}
B.Feigin and F.Malikov, Fusion algebra at a rational level and cohomology of
nilpotent subalgebras of $\2$, hep-th/9310003.

\bibitem{FZ}
V.A.Fateev and A.B.Zamolodchikov, {\it Sov.J.Nucl.Phys.}{\bf 43} (1986) 657.

\bibitem{Aetal}
O.Aharony, O.Ganor, J.Sonnenschein, S.Yankielowicz and N.Sochen,
\NP{399}{1993} 527; \\
H.Hu and M.Yu, On the Equivalence of Non-critical Strings and $G_k/G_k$
Topological Field Theories, Preprint AS-ITP-92-23; \NP{391}{1993} 389.

\bibitem{PG}
P.Furlan, A.Ch.Ganchev, R.Panov and V.B.Petkova, \PL{267}{1991} 63;
\NP{394}{1993} 665.

\bibitem{KK}
V.G.Kac and D.A.Kazhdan, {\it Adv.Math.} {\bf 34} (1979) 97.

\bibitem{KW}
V.G.Kac and M.Wakimoto, {\it Proc.Natl.Acad.Sci.USA} {\bf 85} (1988) 4956.

\bibitem{A}
O.Andreev, \PL{363}{1995} 166.

\bibitem{MO}
N.Marcus and Y.Oz, \NP{392}{1993} 281

\bibitem{DF}
Vl.S.Dotsenko and V.A.Fateev, \PL{157}{1985} 291.

\bibitem{DKetal}
P.Di Francesco and D.Kutasov, \PL{261}{1991} 385;
\newline M.Goulian and M.Li,
{\it Phys.Rev.Lett.} {\bf 66} (1991) 2051;
\newline
Vl.S.Dotsenko, \MPL{6}{1991} 3601.

\bibitem{S}
A.S.Schwarz, Private communication.

\bibitem{BK}
A.Bilal and I.Kogan, Gravitationally dressed conformal field
theory and emergence of logarithmic operators, hep-th/9407151;
\NP{449}{1995} 569.

\end{thebibliography}
\end{document}